\documentclass[useAMS,usenatbib]{mn2e}

\def \kms{{\,km\,s$^{-1}$}}

\def \gcmsq{\hbox{g cm$^{-2}$}}
\def \cmsqg{\hbox{cm$^{2}$ g$^{-1}$}}
\def \arcmin {\hbox{$^\prime$}}
\def \arcsec {\hbox{$^{\prime\prime}$}}

\def\spose#1{\hbox to 0pt{#1\hss}}
\def\ltsim{$\mathrel{\spose{\lower 3pt\hbox{$\sim$}}
        \raise 2.0pt\hbox{$<$}}$\thinspace}
\def\gtsim{$\mathrel{\spose{\lower 3pt\hbox{$\sim$}}
        \raise 2.0pt\hbox{$>$}}$\thinspace}
\def \msun {${\rm M_\odot}$}

\newcommand\solar{\hbox{{$Z_{\odot}$}}}

\newcommand{\apec}{APEC}

\newcommand{\chandra }{{\em Chandra}}

\newcommand{\xspec }{{\em Xspec}}

\newcommand{\ciao }{{\em CIAO}}

\newcommand{\fxunits}{\mbox{ergs cm$^{-2}$ s$^{-1}$}}

\newcommand{\xmm }{{\em XMM}}

\newcommand\omegam{\hbox{{$\Omega_{\rm m}$}}}
\newcommand\omegalambda{\hbox{{$\Omega_{\Lambda}$}}}

\newcommand\ho{\hbox{{$H_{0}$}}}

\usepackage{graphicx}
\usepackage{amssymb}
\usepackage{journal_shortcuts}

\title[The Bullet Group]{Dark matter-baryons separation at the lowest mass scale: the Bullet Group\thanks{Based on observations obtained with XMM-Newton, an ESA science mission with instruments and contributions directly funded by ESA Member States and NASA.}}

\author[Gastaldello et al.]{F. Gastaldello$^{1,2}$\thanks{E-mail: gasta@lambrate.inaf.it}, M. Limousin$^{3,4}$, G. Fo\"{e}x$^{5}$, R. P. Mu\~{n}oz$^{6}$, T. Verdugo$^{7}$, V. Motta$^{5}$, \newauthor A. More$^{8,9}$, R. Cabanac$^{10}$, D. A. Buote$^{2}$, D. Eckert$^{11,1}$, S. Ettori$^{12,13}$, A. Fritz$^{1}$, \newauthor S. Ghizzardi$^{1}$, P. J. Humphrey$^{2}$, M. Meneghetti$^{12,13,14}$, M. Rossetti$^{15,1}$\\
$^{1}$INAF - IASF Milano, via E. Bassini 15, I-20133 Milano, Italy.\\
$^{2}$Department of Physics and Astronomy, University of California at Irvine, 4129 Frederick Reines Hall, Irvine, CA 92697-4575, USA\\
$^{3}$Aix Marseille Universit, CNRS, LAM (Laboratoire d~Astrophysique de Marseille), UMR 7326, 13388, Marseille, France\\
$^{4}$Dark Cosmology Centre, Niels Bohr Institute, University of Copenhagen, Juliane Maries Vej 30, DK-2100 Copenhagen, Denmark\\
$^{5}$Instituto de F\'{i}sica y Astronom\'{i}a, Universidad de Valpara\'{i}so, Avda. Gran Breta\~{n}a 1111, Valpara\~{i}so, Chile\\
$^{6}$Instituto de Astrof\'isica, Facultad de F\'isica, Pontificia Universidad Cat\'olica de Chile, Av. Mackenna 4860, 7820436 Macul, Santiago, Chile\\ 
$^{7}$Centro de Investigaciones de Astronom\'{i}a, AP 264, M\'{e}rida 5101-A, Venezuela\\ 
$^{8}$Kavli Institute for Cosmological Physics, U. of Chicago, 5640 S. Ellis Ave., Chicago IL-60637, USA\\
$^{9}$Kavli IPMU, U. of Tokyo, 5-1-5 Kashiwanoha, Kashiwa, 277-8583, Japan\\  
$^{10}$Universite de Toulouse-UPS, CNRS; Institut de Recherche en Astrophysique et Plan´etologie; 57 avenue d~Azereix, 65000 Tarbes, France\\
$^{11}$Astronomical Observatory of the University of Geneva, ch. d~Ecogia 16, 1290 Versoix, Switzerland\\
$^{12}$INAF, Osservatorio Astronomico di Bologna, via Ranzani 1, I-40127, Bologna, Italy\\ 
$^{13}$INFN, Sezione di Bologna, viale Berti Pichat 6/2, I-40127, Bologna, Italy\\
$^{14}$JPL, 4800 Oak Grove Dr., Pasadena, CA 91109, USA\\
$^{15}$Universit\'a degli studi di Milano, Dip. di Fisica, via Celoria 16, 20133 Milano, Italy\\}
\date{\today}

\pagerange{\pageref{firstpage}--\pageref{lastpage}} \pubyear{2014}

\begin{document}

\label{firstpage}

\maketitle

\begin{abstract}
We report on the X-ray observation of a strong lensing selected group, SL2S J08544-0121, with a total mass of $2.4\pm 0.6 \times 10^{14}$ \msun\ which 
revealed a separation of $124\pm20$ kpc between the X-ray emitting collisional gas and the collisionless galaxies and dark matter (DM), traced by 
strong lensing. This source allows to put an order of magnitude estimate to the upper limit to the interaction cross section of DM of 
10 cm$^{2}$ g$^{-1}$. It is the lowest mass object found to date showing a DM-baryons separation  and it reveals that 
the detection of bullet-like objects is not rare and confined to mergers of massive objects opening the possibility of a statistical detection of 
DM-baryons separation with future surveys.
\end{abstract}


\section{Introduction}

Merging galaxy clusters are unique astrophysical probes of the properties of dark matter (DM), which accounts for the
majority of the mass in the universe. During a cluster merger, the cluster galaxies are collisionless particles, 
affected only by gravitational interactions, while the X-ray emitting plasma clouds, the dominant baryonic 
components in mass, are slowed down by ram pressure. Collisionless DM behaves as the galaxies
so as the merging progresses the DM component is separated from the X-ray gas \citep{Furlanetto.ea:02}. 
The presence of DM and constraints on its self-interaction cross section can therefore be inferred by
measuring a spatial offset between the X-ray emission of the plasma and its total mass distribution as revealed by gravitational
lensing, which is independent of the type of matter present.

An offset between the X-ray gas distribution and the mass inferred from gravitational lensing was detected 
for the first time in the Bullet cluster \citep[][M04 hereafter]{Markevitch.ea:04}. A few other examples of 'bullet-like 
clusters' have been found following that discovery 
\citep[e.g.,][]{Bradac.ea:08,Merten.ea:11,Dawson.ea:12,Dahle.ea:13}.
Since the collisions between two massive progenitors are rare \citep{Shan.ea:10} the number of detected massive clusters 
undergoing mergers with the proper configuration is not expected to increase significantly.
The utility of a small number of individual systems is limited by observational uncertainties in their collision
velocity, impact parameter and angle with respect to the plane of the sky \citep[e.g.,][]{Dawson:13}. Here we show
that these studies can be extended to less massive systems which are much more numerous than massive clusters.

All distance-dependent quantities have been computed assuming a cosmological model with \omegam=0.3, \omegalambda=0.7 and 
\ho=70 km s$^{-1}$ Mpc$^{-1}$.   
\begin{figure}
\begin{center}
\includegraphics[width=9.0cm]{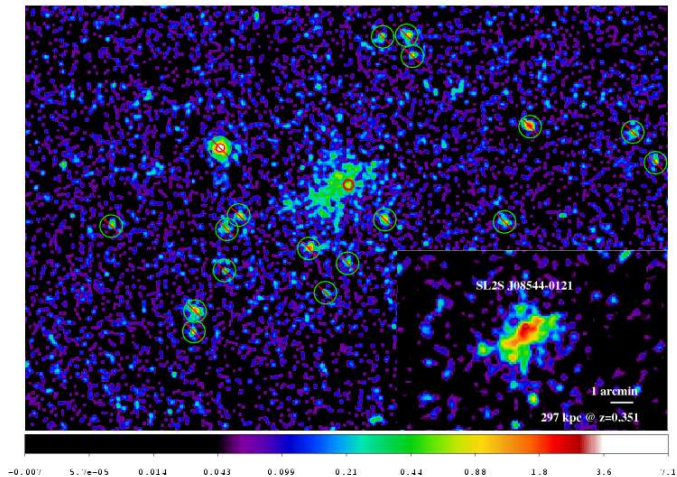}
\caption{The EPIC 0.5-2 keV exposure-corrected 
and particle background subtracted image of SL2S\,J08544. The detected point sources are highlighted by the green
circles; with the red circle of radius 11\arcsec\ the point source within the group diffuse emission
is also indicated. Point sources have been removed with the \ciao\ tool {\em dmfilth} which replaces photons within the source
with a locally estimated background. This processed image of the diffuse emission is shown in the inset in the bottom right corner
and it has been used to produce the X-ray contours shown in Fig.\ref{fig.bullet}. The image has been smoothed with a 7.5\arcsec\ Gaussian 
(15\arcsec\ for the inset). 
The color bar is in units of counts per pixel.}
\label{fig.xray}
\end{center}
\end{figure}

\section{The object SL2S J08544-0121}\label{sect_obj}

The object SL2S J08544-0121 is a gravitational lens found in the Strong Lensing
Legacy Survey (SL2S), a semi-automated search for strong lensing systems on the Canada-France-Hawaii Telescope 
Legacy Survey (CFHTLS) Deep and Wide fields \citep{Cabanac.ea:07}. SL2S uses an algorithm aimed at detecting 
efficiently group scale lenses, with image separations of the order 3\arcsec-12\arcsec, intermediate 
between the ones found in galaxies (1\arcsec-2\arcsec) and massive 
clusters (\gtsim 20\arcsec) \citep{More.ea:12}. We will therefore define SL2S J08544-0121 as a group based on this criterion.
SL2S\,J08544 is located at redshift 0.35 and it displays a bimodal light distribution
with a strong lensing system located at one of the two luminosity peaks separated
by $54\arcsec$ (267\,kpc transverse physical distance). The strong lensing features detected from ground based images 
have been followed up with ACS camera on the \emph{Hubble Space Telescope} (HST), revealing a 
very perturbed lensing configuration. Indeed, the main arc
and the counter-image of the strong lensing system are located at $\sim 5\arcsec$ and
$\sim 8\arcsec$ from the lens galaxy center respectively. It was found \citep{Limousin.ea:10} 
that a simple elliptical isothermal potential centered on the lensing galaxy
could not satisfactorily reproduce the strong lensing
observations. One straightforward way to measure the lensing model quality  is to quote the RMS error which quantifies the distance 
between the observed position of the lensed images and the position derived from the lensing mass model.
The smaller this distance, the better the mass model. For this model the RMS error in the image position is 0.38\arcsec. It was found that 
one needs to take into account in the modeling the mass distribution of the galaxy group within which the lens is embedded. In addition to the mass
associated with the strong lensing deflector, mass associated with the second
luminosity peak is required in order to accurately reproduce the strong lensing observations (RMS of 0.05\arcsec), 
demonstrating that SL2S\,J08544 displays a bimodal mass distribution following the
light distribution. 
If on the contrary we construct a mass model where we add a mass clump consistent with the X-ray gas distribution (see Section \ref{sect_xray})
we are not able to improve the fit (RMS of 0.36\arcsec) with respect to the unsuccessful model where a single mass clump is associated with the SL 
deflector.
The total mass of the group inferred from the strong lensing analysis is found to be in
good agreement with an independent weak lensing analysis \citep{Foex.ea:13}.
Spectroscopic follow-up of 18 ellipticals with FORS\,2 at VLT confirms the presence of a
galaxy group at $z=0.35$ \citep{Munoz.ea:13}. In particular, the two brightest galaxies populating the two
luminosity peaks are found at the same redshift. We do not confirm the bimodal distribution of velocities
suggested in \citet{Munoz.ea:13}: the redshift histogram has indeed a broad high velocity tail but a series
of tests looking for departure from Gaussianity returned a null result. We applied a Anderson-Darling test
\citep[e.g.,][]{Hou.ea:09} which returned a $p$ value of 0.1553, therefore consistent with a Gaussian distribution. We also applied to the data
the Kaye's Mixture Model (KMM) test \citep{Ashman.ea:94} which returned a $p$ value of 0.06 for a two-group partition over a single group.
The number of spectroscopic members does not allow a more detailed view of the merger dynamics and in particular
the component of the velocity along the line of sight.

\section{The XMM observation of SL2S J08544-0121}\label{sect_xray}

We observed SL2S J08544-0121 with \xmm\ as part of an X-ray follow-up program of the SL2S groups 
to obtain an X-ray detection of these strong-lensing selected systems and a measurement of the X-ray luminosity and 
temperature. 
SL2S J08544-0121 was observed by \xmm\ for 9.5 ks for the MOS detector and 5 ks for the pn detector.
The data were reduced with SAS v12.0.0 using the tasks
{\em emchain} and {\em epchain}. We considered only event patterns
0-12 for MOS and 0 for pn and the data were cleaned using the standard
procedures for bright pixels and hot columns removal and pn out-of-time correction.
Periods of high backgrounds due to soft protons were filtered out but their impact was negligible for this observation.
We checked the observation for contamination by solar wind charge exchange: 
ACE (Advanced Composition Explorer) SWICS O+7/O+6 ratio was less than 0.3, a value which is typical of
the quiescent Sun \citep{Snowden.ea:08}. No variation in the light curve in the soft (0.5-2 keV) energy band was detected
as a further check of negligible contamination by this background component.

For each detector we created images in the 0.5-2 keV band with point sources,
detected using the task {\em edetect\_chain}, masked using circular
regions of 25\arcsec\ radius centered at the source position. 
Point-source-free, cleaned images have been generated with the \ciao\ tool {\em dmfilth} which replaces photons within the source
with a locally estimated Poisson-deviated noise. 
The images have been exposure corrected and background subtracted using the XMM-Extended Source
Analysis Software (ESAS).
The \xmm\ image in the 0.5-2 keV band of the field of SL2S J08544-0121 is shown in Fig.\ref{fig.xray}.
The X-ray emission of SL2S J08544-0121 is clearly extended: the best-fit $\beta$-model to the surface brightness
profile has a core radius of $r_c = 128^{+64}_{-49}$ kpc ($26\arcsec^{+13}_{-10}$) and $\beta=0.52^{+0.09}_{-0.06}$.
The faint point source (a 3$\sigma$ detection) in the SW embedded within the extended emission of the group has been replaced
within a circle of 11\arcsec\ radius with a locally estimated Poisson noise with the same procedure adopted for the other point sources 
in the image. This is the largest region not overlapping with the peak of the emission itself (and corresponding
roughly to 60\% of the encircled energy fraction, EEF, for a point source on-axis). We estimate that the
possible contamination by the emission in the wings of the PSF corresponds to 1\% of the extended emission.
The \xmm\ astrometry is known to be accurate to within 1\arcsec\ \citep{Guainazzi.ea:13} and we quantified the error
in the determination of the peak by calculating the error in the position of the center of 
a two-dimensional beta model fitted in a region of 30\arcsec\ radius (80\% of the EEF) 
with the CIAO software {\em Sherpa} \citep{Freeman.ea:01} which is of the order 3\arcsec.
We therefore estimate the error in the position of the X-ray peak to be 4\arcsec. 
If instead of replacing the weak point source we model it with an appropriate PSF model we obtain
the same position for the X-ray peak within the errors.

To assess if the possible systematic error due to undetected source might be larger than the above estimate, 
we tested our analysis on simulations of $10^{3}$ \xmm\
images with a source list with flux distribution and source density computed using the Log ($N$) - Log($S$)
from \citet{Moretti.ea:03} down to a level of $1 \times 10^{-17}$ \fxunits\ and we then added the extended emission of the source; 
the main instrumental characteristics (PSF, vignetting, background) were taken into account. The standard deviation
of the distribution of position of peaks was 3\arcsec, consistent with the estimated error.

\begin{figure}
\begin{center}
\includegraphics[width=5.5cm,angle=-90]{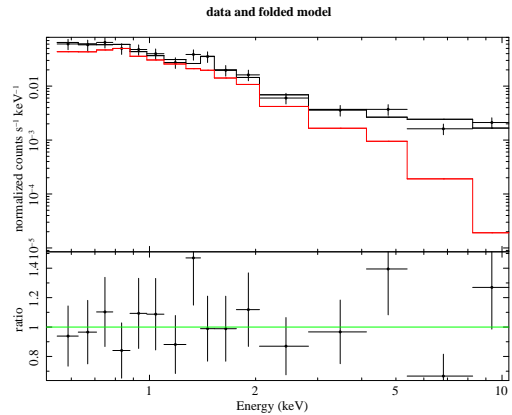}
\caption{pn spectrum extracted from a 1.5\arcmin\ aperture:
all the background components, instrumental and cosmic, have been modeled and the source component is shown by
the solid red line. The ratio of data over the model are also shown.}
\label{fig.spectra}
\end{center}
\end{figure}

\begin{figure*}
\begin{center}
\includegraphics[width=0.83\hsize,clip]{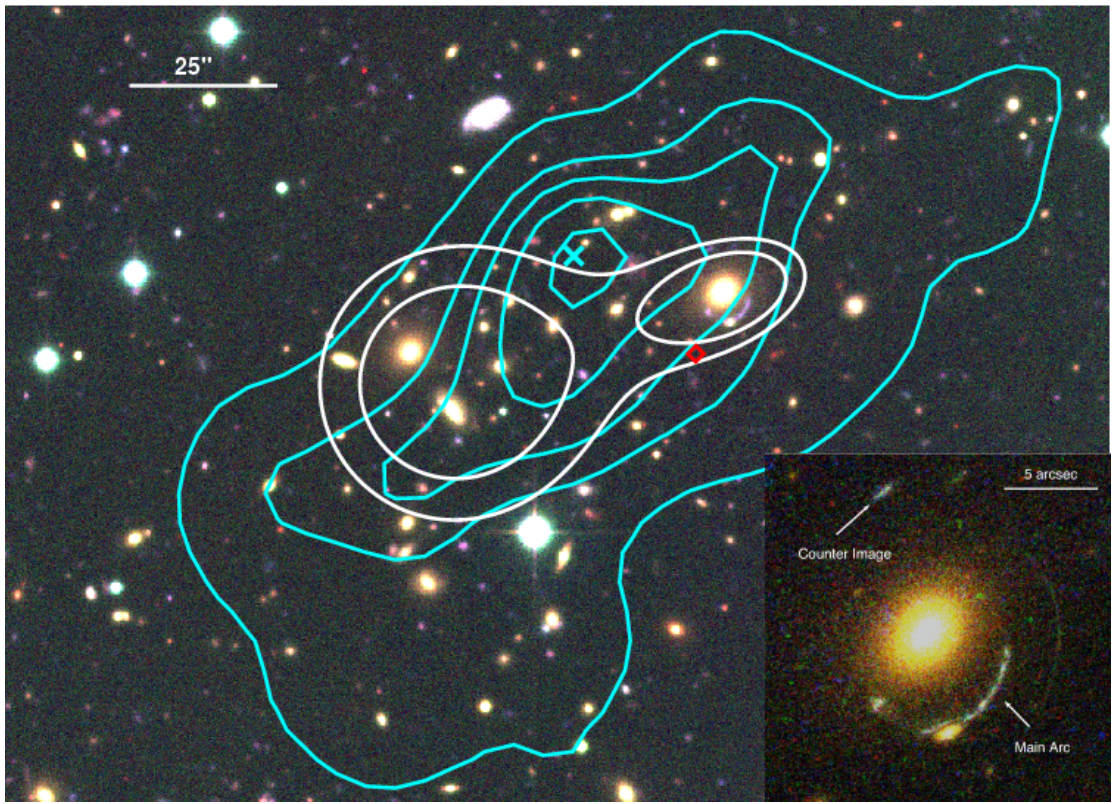}
\caption{Composite CFHTLS \emph{g},\emph{r},\emph{i} color image of the group SL2S J08544-0121, with a size of
3\arcmin\ $\times$ 2\arcmin\ corresponding to 891 kpc $\times$ 594 kpc at the redshift of the object, z=0.351. 
Overlayed in white are the mass contours derived from the strong lensing model, showing where the projected mass density 
equals $(1.5, 2.0) \times 10^{10}$ \msun\  arcsec$^{-2}$. 
The X-ray contours from the \xmm\ observation are over-plotted in cyan, showing the clear displacement between the X-ray peak, marked by the cyan x point, and the lensing mass, centered at the coordinates of the galaxy populating the lens (see insert). The red diamond point marks the position of the
excluded point source discussed in the text. The insert in the bottom right corner
shows the HST image with the main arc and its counter image, revealing the particular asymmetric configuration. 
}
\label{fig.bullet}
\end{center}
\end{figure*}

For spectral fitting, we extracted spectra for each detector from a 1.5\arcmin\ aperture
centered on the peak of the emission, with radius chosen to maximize the S/N over the
background. Redistribution matrix files (RMFs) and
ancillary response files (ARFs) were generated using the SAS tasks
{\em rmfgen} and {\em arfgen} in extended source
mode. 
The spectra from the three detectors were
jointly fitted with an \apec\ thermal plasma \citep{Smith.ea:01} modified by Galactic
absorption \citep{Kalberla.ea:05}. The spectral fitting was performed with
\xspec\ \citep{Arnaud:96} in the 0.5-12 keV band (0.5-13 keV for the pn) using the C-statistic and quoted
metalicities are relative to the abundances of \citet{Grevesse.ea:98}. 
To account for the background we included additional spectral components in the fits: we included two \apec\ components 
(kT = 0.07 keV and 0.2 keV, the former un-absorbed) to account for the Galactic foreground and a power law component ($\Gamma = 1.41$) for the
Cosmic X-ray Background due to unresolved AGNs. To account for the instrumental background, we included 
a number of Gaussian lines and a broken power law models, which were not folded through the ARF. The background parameters were constrained
by fitting spectra extracted from an annular region close to the source extraction region and in a larger annulus 
of 10\arcmin-12\arcmin\ and then the fitted normalizations were rescaled accordingly to the source extraction area. 
The estimated 0.5-2 keV background is in good agreement with the ROSAT All-sky Survey (RASS) spectrum obtained 
using the X-ray Background Tool \citep{Snowden.ea:97}.
We obtained a fit with a C stat/dof value of 31/36 and the best fit parameters are kT = $3.5^{+0.6}_{-0.5}$ keV and 
$Z=0.6^{+0.7}_{-0.5}$ \solar. We show in Fig.\ref{fig.spectra} the pn spectrum and the best fit model, highlighting
the source component.
As a further check to the maximum likelihood fitting we use Markov chain Monte Carlo (MCMC) techniques and Bayesian inference
in \xspec\ to constrain the confidence level of the temperature determination. 
We allowed for a 10\% systematic in our background modeling setting Gaussian priors on the rescaled 
normalizations of the cosmic and particle background with a width of 10\% of their best fit values and we set constant priors
on the source parameters. We produced a chain of length $10^{4}$ steps, after an initial ``burn-in'' of 5000 steps, with
the Metropolis-Hastings sampler. We then marginalized over the all other parameters to generate a posterior 
probability for the temperature. The 68\% confidence interval is (3.0, 4.5) keV.

For the total mass determination of the system we used the chain outputs for the temperature parameter as inputs in the M-T scaling relation 
of \citet{Arnaud.ea:05} obtained for the full sample investigated in that paper and within an over-density of 200: 
we find an estimate for the mass of the system of $\rm{M_{200}}= 2.4\pm0.6\,\times 10^{14}$ \msun\ which is in good agreement with the lensing 
determination of $\rm{M_{200}}= 2.2^{+0.4}_{-0.6}\,\times 10^{14}$ \msun, derived adopting the value of dispersion velocity, 
$644^{+69}_{-102}$ \kms, of the Single Isothermal Sphere model used in the modeling of the weak lensing data \citep{Foex.ea:13}. Simulations indicate that clusters even in the middle of a major merger follow the M-T relation with a scatter of about 20\%-25\% 
along the mass axis \citep[e.g.,][]{Rasia.ea:11}.

\section{DM-baryons separation}

The X-ray data reveal a single peak located between the two
light/mass clumps with a clearly extended and disturbed morphology and an elongation of the emission along
the South-East direction. These features are typical of an advanced merger event 
with the two clumps having already experienced their first passage following a South-East North-West direction.
The geometry of an elongated single X-ray feature between the two DM clumps revealed by lensing is analogous to the 
configuration seen in the cluster MACS J0025.4-1222 \citep{Bradac.ea:08}. The peak of the baryonic mass 
distribution derived from the X-ray emission is offset by 25\arcsec$\pm$4\arcsec\  
from the strong lensing system in the North-West, corresponding to a transverse physical separation of 
$124\pm20$ kpc at the redshift of the object and given our adopted cosmology (see Fig.\ref{fig.bullet}.)

The mass estimate of $\rm{M_{200}}= 2.4\pm0.6\,\times 10^{14}$ \msun\
makes SL2S J08544-0121 the smallest mass system found to date for which a 
significant DM-baryons separation has been detected, $\sim 7$ times less massive than the Bullet cluster
and $\sim$ 2 times less massive than the "Burst" cluster, ZWCl 1234.0+02916 \citep{Dahle.ea:13}, the currently known 
smallest mass system showing a bullet-cluster like configuration.

Using the data of SL2S J08544-0121 we can derive another independent order-of-magnitude estimate for the 
self-interaction cross section of DM. We follow the first argument of M04
and assume that the NW subcluster has passed once close to the center of the SE subcluster and that the direction
of motion is very nearly in the plane of the sky. The offset between the DM centroid of the NW subcluster and the
peak of the X-ray gas, which is assumed to belong originally to the same 
structure, indicates that the scattering depth of the DM subcluster with respect to the collisions of the DM particle stream is lower than 
1. If this would not be the case, the DM would behave as a fluid showing ram pressure stripping as the gas.
We further assume that the surface mass density along the collision direction is similar to that along the line of sight: this is a conservative estimate 
given the best fit ellipticity of the halo from strong lensing \citep[$0.50\pm0.04$,][]{Limousin.ea:10}.
The scattering depth of DM in the strong lensing subcluster is therefore

\begin{equation}
\tau  = \frac{\sigma}{m}\Sigma < 1
\label{bspind}
\end{equation}
where $\sigma$ is the DM collision cross section, $m$ is its particle mass, and $\Sigma$ is the 
DM surface density of the NW subcluster associated to the lens.
The choice of the radius where to calculate the average DM surface density in previous studies has been selected to provide conservative 
constraints and a comparison between different cluster systems. A radius of 150 kpc was used for the subcluster 
of the Bullet cluster given the available lensing data at that time (M04) and then used as a reference in following studies. 
If we calculate $\Sigma$ at the position of the strong lensing peak averaged over a radius of 150 kpc we obtain 
$0.06$ \gcmsq\ and therefore an estimate for the upper limit on the scattering cross section of $\sigma/m$ \ltsim\ 17 \cmsqg.
If instead we use a radius smaller than the measured separation of 124 kpc, in particular 100 kpc, we get $0.10$ \gcmsq\ and an 
upper limit of $\sigma/m$\ltsim\ 10 \cmsqg.

The detection of DM-baryons separation in SL2S J08544-0121 provides further evidence for the collisionless DM 
model and an independent upper limit on its interaction cross section. 
Systems with mass of $1-2 \times 10^{14}$ \msun\ like the Bullet Group are $10^3$ times more numerous than massive clusters of 
$1 \times 10^{15}$ \msun\ like the Bullet Cluster and therefore examples of DM-baryons separation are not as rare as usually assumed 
\citep{Amendola.ea:13}. Indeed numerical simulations already 
suggested a fair number of ``bullets'' at every mass scale \citep[e.g.][]{Forero-Romero.ea:10}, in particular the number of bullet groups is three times 
larger than the one of bullet clusters \citep{Fernandez-Trincado.ea:14}.
Upcoming lensing surveys (e.g. with the \emph{Euclid} satellite) and X-ray surveys (with the \emph{eROSITA} 
telescope on the Spektrum-Roentgen-Gamma Mission) should therefore provide hundreds of similar examples 
allowing the properties of DM to be studied in a statistical manner \citep{Massey.ea:11}.

\section{Conclusions}

SL2S J08544-0121 is a really remarkable object because it is a strong lensing selected group, its
lensing image configuration already provided evidence of a bimodal mass configuration and the X-ray follow-up
showed evidence of baryons-DM separation down to at least a mass scale of few times $10^{14}$ \msun.
It might be considered as a proof of concept of the potential of upcoming deep lensing and X-ray surveys
of discovering many similar examples allowing a statistical study of the properties of DM.
Deeper X-ray observations with \chandra\ and optical spectroscopy of an increased number of member galaxies
are needed for a better understanding of the merger geometry of this system.

\section*{Acknowledgments}
FG is supported by INAF-ASI through grant I/023/05/0, I/088/06/0 and I/032/10/0. 
AF is supported by INAF VIPERS PRIN 2008/2010. V.M. is supported by FONDECYT 1120741.
R.M. is supported by FONDECYT 3130750. 
R.C., G.F., M.L., V.M. is supported by ECOS-CONICYT C12U02. 
TV is supported by CONACYT through grant 165365 and 203489.
MM is supported by ASI/INAF/ I/023/12/0 and from INFN project PD51.

\bibliography{gasta}

\bibliographystyle{mn2e}

\label{lastpage}

\end{document}